\documentclass[preprintnumbers,a4paper, amsmath, amssymb, floatfix,
nofootinbib, twocolumn, wrapfloat byrevtex]{revtex4}
 \usepackage{amsmath,mathrsfs,graphicx,epstopdf,placeins,float}
 \usepackage{booktabs}
 \usepackage{multirow}
 \usepackage{pstricks}
 \usepackage{epsfig}
 \usepackage{bm}

 \newcommand{\beq}[1]{\begin{equation}\label{#1}}
 \newcommand{\eeq}{\end{equation}}
 \newcommand{\bea}[1]{\begin{eqnarray}\label{#1}}
 \newcommand{\eea}{\end{eqnarray}}

\begin{document} 
\title{Spin-Dependent Scattering of Scalar and Vector Dark Matter on the Electron}
\author{Ke-Yun Wu}
\email{keyunwu@emails.bjut.edu.cn}
\affiliation{Faculty of Science, Beijing University of Technology, Beijing 100124, China}
\author{Zhao-Hua Xiong}
\email{xiongzh@bjut.edu.cn}
\affiliation{Faculty of Science, Beijing University of Technology, Beijing 100124, China}
 \begin{abstract}
 The property of dark matter is unknown so far. However, a model-independent classification of dark matter candidates can be achieved by using various symmetries, as done in the Standard Model. Fermionic dark matter has been researched extremely, one favored candidate is the neutralino in the Minimal Supersymmetric Standard Model, which are required by fermion-boson symmetry and preservation of R-parity. Bosonic dark matter has not been studied sufficiently, especially the scenario of dark matter with mass of sub-GeV. In this paper, we consider the effect of spin-dependent (SD) on scalar and vector dark matter, which are mediated by pseudo-scalar and axial-vector, and evaluate effect on the dark matter-electron scattering cross section. We list all the interaction and form factor of dark matter-electron SD scattering, and use XENON10/100/1T experiment data to derive the exclude limit of SD cross section. We find that the SD scattering of scalar and vector dark matter can be three orders of magnitude stronger than spin-independent (SI) scattering, due to the $p$-wave scattering.

 \end{abstract}
\maketitle

\section{Introduction}
	About 27\% of the energy density of our universe consists of dark matter which is non-luminous and rarely interacts with baryons \cite{Young:2016ala}. Beyond gravitational phenomena, dark matter still has many elusive properties which include its mass and interactions, etc. Usually the methods of exploration for the dark matter (DM) include direct detection, indirect detection and collider experiments. Among all the candidates of dark matter, the most attractive one may be the Weakly Interacting Massive
	Particles (WIMPs). Such this type candidates for dark matters is the lightest Majorana fermion, 1ie., neutralino in the Minimal Supersymmetric Standard Model, which are required by fermion-boson symmetry and preservation of the global symmetry called R-parity. Clearly, the prospects for direct detection of dark matter depend crucially on whether WIMP-nucleon interactions are primarily spin-dependent or spin-independent. Although property of dark matter is known so far, one can perform a model-independent classification of dark matter candidates using symmetries. Firstly, WIMP dark matter is constrained to be neutral under both electromagnetism and color; secondly, if the WIMP-nucleon cross section is primarily
	spin-dependent, scalar dark matter is disfavored; thirdly, the DM-nucleus interaction is described by an effective potential, which is a rotationally invariant scalar formed out of vectors: the relative velocity, DM position, and nuclear spin \cite{Fan:2010gt}. By using the effective potential and possible form of the Lagrange, we can evolute the leading contributions to the cross section.
	
	Previous studies focus mainly on DM with mass at the order of 100 GeV \cite{Dodelson:2003ft}. However, the recent dark matter detection, the limits of WIMP cross section approach the ``neutrino floor" more and more closer, less parameter space are left for the detection. Instead, there still are large space for the detection
	for the light dark matter with mass lighter than $1\rm GeV$, as for the much smaller momentum transfer in the collision. Such light dark matter can scatter with electrons, causing ionization of atoms in a detector target material and leading to single- or few-electron
	events. We use 15 kg-days of data acquired in 2006 to set limits on the dark matter-electron scattering cross section. As for the detection of the sub-GeV dark matter, the ionization signal will become more promising, this is because of that the electron recoil energies can exceed the threshold of direct detection when the sub-GeV dark matter scattering on the electron \cite{Essig:2011nj,Essig:2015cda,Athron:2020maw,Gao:2020wer,An:2020bxd,Ge:2020jfn,Du:2020ybt,Choi:2020ysq,Bloch:2020uzh,Knapen:2021run,Chao:2021liw,Hochberg:2021pkt,Essig:2012yx,Essig:2017kqs,Dolan:2017xbu,Baxter:2019pnz,Su:2020zny}. Especially, the dual-phase time projection chamber (TPC), liquid XENON detector has given the most stringent constraint on the detection limits. When the dark matter particles scatter with the electron, the ``S2" signal of electron can be produced by electric fields upward into gaseous xenon in the liquid. The resulting recoil electrons have sufficient energy to ionize other atoms. The resulting signal includes the observed electron and scintillation photon. In general, the experiment data are adopted from XENON10, XENON100 and XENON1T~\cite{XENON100:2011cza, XENON100:2013ele, XENON:2016jmt,XENON:2019gfn}, respectively.
	
	Usually the scattering of dark matter are divided into the SI and SD case. The SI scattering, which is originated form the scalar couplings between dark matter and the target, dominate in the detectors with the heavy nucleus. While the SD scattering, which is originated form the spin couplings, dominate in the detectors with the light nucleus \cite{Agrawal:2010fh,Freytsis:2010ne,Fitzpatrick:2012ib,Han:2016qtc,Abdughani:2017dqs,Wang:2021nbf}. For SI scattering, the interaction will be mediated by a scalar or vector. The SD scattering can be modeled by setting the dark matter interacts solely with electron spin with a pseudo-scalar or axial-vector mediator~\cite{Ramani:2019jam}. Then $p$ wave scattering always happens in the SD scattering, and the exclusion limit of the corresponding cross section can be enhanced by the momentum transfer \cite{Liu:2021avx}. This is a very different point compared with the SI scattering. And SD limits of the fermionic dark matter have bee thoroughly studied in the literature \cite{Ramani:2019jam,Agrawal:2010fh}.
	However the SD scattering of the scalar and vector dark matters still needs deeper studies, especially the effects of the momentum transfer by the mediator.
	In this work, scalar and vector types of dark matter are proposed and the corresponding effective operators are studied. The SD scattering of these dark matter, especially the effects of the momentum transfer, are studied numerically. In SD scattering case, the spin of electron will make the response function and form factors more complicated than the simple scalar interaction. Therefore, these form factor and response function are calculated and the corresponding the exclusion limit on the SD cross sections are discussed.
	
	The paper is organized as follows: the theoretical calculation of cross section and the benchmark model of scalar and vector dark matter are shown in the Sec.~\ref{sec2}. In Sec.~\ref{sec3}, we described the recoil spectra and numerical results. The conclusion
	are given in Sec.~\ref{sec4}.

\section{The dark matter model and the form factor}\label{sec2}
\subsection{The calculation framework 
of dark matter and electron scattering process}
The scattering between dark matter and the electrons are calculated under 
the impulse approximation in which the binding of the electrons are neglected.
As talked above, the detailed calculations of the scattering are divided into 
the SI and SD cases.
In case of the detection of the recoils of the nuclei,
the rate of the SI case is proportional to the atom number 
thus it is sensitive the heavy nuclei. 
However, the spin of nuclei is always considered as 
coming from the single last unpaired nucleon, then
the detection will be sensitive the light nuclei, relatively.
The calculation of the SD scattering on the nuclei are different from the SI case.
For the scattering on the electron, both SI and SD cases are scattering on a single electron. Thus at the first glance,
all the scatterings should be taken into account as they seem
to have the same weights from theoretical side.
What we can do in the simplest way is to consider couplings on 
 the electron operators
 \begin{eqnarray}
e^{\dagger}{\bm I }e,~~e^{\dagger}{\bm S}e.
\end{eqnarray}
The couplings to the first one can be seen as to cause a SI scattering,
while the couplings to the second one will cause a SD scattering. 
The spin operator $\bm S$ is the non-relativistic limit of the quantum fermionic
field which always comes from the $\gamma$ matrices then from the 
Dirac spin 
\begin{eqnarray}
  {\bm \sigma}^{s}=2{\bm S},
\end{eqnarray}
with 
\begin{eqnarray}
  {\bm \sigma}^{s}=\gamma^{0}{\bm \gamma}=
\begin{pmatrix}{\bm \sigma} & 0\\0 & {\bm \sigma} \end{pmatrix}.
\end{eqnarray}
In general, the spin operators originate from a pseudo-scalar or a axial-vector mediator.
Both SI and SD scattering can be considered in a uniform formula which will be talked in the following

In the direct detection, the measured quantity is the counting rate $R^{\rm ion}$ which 
is theoretically predicted by the velocity-averaged cross section between dark matter
and the target. In direct detection, the differential event rate that is induced by ionizing electron can be written as 
\begin{eqnarray}
\label{rion}
 \frac{d R^{\rm ion}}{d E_{R}}
&=&
N\frac{\rho_{\chi}}{m_{\chi}}\frac{\bar{\sigma}_{e}}{8\mu_{\chi e}^{2}}\\
&& \times \int qdq \vert F_{\rm DM}(q)\vert^{2}
{\cal R}_{(\rm ion)}(E_R,q)\eta(v_{\rm min}),\nonumber
\end{eqnarray} 
in which, $\rho_{\chi}=0.4~\rm GeV/cm^{3}$ is local dark matter density and the number of the target atom is $N$. $E_{R}$ and $q$ are the recoil energy and momentum transfer of the electron, respectively. $\bar\sigma_e$ is the reduced cross sections and $\mu_{\chi e}=m_{\chi}m_{e}/(m_{\chi}+m_{e})$ is the reduced mass of dark matter ($m_\chi$) and electron ($m_e$).

 In Eq. \eqref{rion}, we apply the response function ${\cal R}_{(\rm ion)}$ to denote the wave function of electron ionization $\vert f_{\rm ion}^{nl}\vert^{2}$. Compared with SI scattering, the interaction in SD scattering will consider the spin operator of electron which comes from the Dirac spin $\boldsymbol{\sigma}^{s}$. Thus in this case, the wave function of electron is not plane wave $e^{i{\bm{q}}\cdot {\bm {r}}}$ but is a more complex form $e^{i{\bm{q}}\cdot {\bm {r}}}{\bm {\sigma}^{s}}$. We follow the Ref. \cite{Liu:2021avx} to write these two functions as 
 \begin{eqnarray}
 {\cal R}_{\rm SI}^{\rm ion}(E_R,q)&=&\sum_{J_{f}}\overline{\sum_{J_{i}}}
\left\vert\langle J_{f}\vert\sum_{i=1}^{Z}e^{i{\bm q}\cdot{\bm r}_{i}}\vert  
J_{i}\rangle\right\vert^{2}\nonumber\\
 && \times\delta(E_{J_{f}}-E_{J_{i}}-E_R),
 \label{eqRSI}
 \end{eqnarray}
 \begin{eqnarray}
 {\cal R}_{\rm SD}^{\rm ion}(E_R, q)&=&\sum_{J_{f}}\overline{\sum_{J_{i}}} \sum_{k}
\left\vert\langle J_{f}\vert\sum_{i=1}^{Z}e^{i{\bm q}\cdot{\bm r}_{i}}{\bm \sigma}_{i,k}^{s} 
\vert J_{i}\rangle\right\vert^{2}\nonumber\\
 &&\times\delta(E_{J_{f}}-E_{J_{i}}-E_R).
 \label{eqR}
 \end{eqnarray}
In which, the $\langle J_{f}\vert\cdots\vert J_{i}\rangle$ is 
the transition amplitude from initial state $\vert J_{i}\rangle$ to the final state $\langle J_{f}\vert$. ${\bm \sigma}_{i,k}^{s}$ is Dirac spin.
The delta function impose the energy conservation of the transition. 
The summations all the electron at different shells gives the response functions. The ratio $r$ of SD and SI response function is given by (see Ref \cite{Liu:2021avx,Chen:2015pha,Catena:2019gfa})
\begin{eqnarray}
 r=\frac{ {\cal R}_{\rm SD}^{\rm ion}}{{\cal R}_{\rm SI}^{\rm ion}}= 3.
\end{eqnarray}
The ratio $r=3$ only an approximate result 
in which the dependence on the recoil energy $E_{ R}$ 
are ignored. In fact, as shown in Ref.~\cite{Liu:2021avx}, 
the ratio depends on the $E_{R}$, of which the detailed calculation is shown in Appendix.

The $\eta(v_{min})$ which is the velocity distribution of dark matter follow the Maxwell-Boltzmann. In the calculation, we assume that the incoming velocity of dark matter is $\boldsymbol v$, and the outgoing momentum of dark matter and electron are $\boldsymbol p_{\chi}'$ and $m_{e}\boldsymbol v_{e}$. Therefore the momentum conservation is given by
   \begin{eqnarray}
       \boldsymbol q=m_{\chi}\boldsymbol v-\boldsymbol p_{\chi}'=m_{e}\boldsymbol v_{e},
   \end{eqnarray}
and the energy conservation is given by
\begin{eqnarray}
 \Delta E_{e}=\frac {1}{2}m_{\chi}v^{2}-\frac{\vert m_{\chi}v-q\vert^{2}}{2m_{\chi}}=\bm{v}\cdot \bm{q}-\frac {\bm{q}^2}{2m_{\chi}},
 \label{energy}
\end{eqnarray}
where $\Delta E_{e}=E_{\rm b}+E_{R}$ is transferred energy which dependent on binding energy $E_{\rm b}$ and recoiled energy $E_{R}$. From Eq. \eqref{energy}, the minimum velocity $v_{\rm min}$ required for ionizing electron in $(n,l)$ shell is given by
\begin{equation}
v_{\rm min}=\frac{|E_{\rm b}^{nl}|+E_{R}}{q}+\frac{q}{2m_{\chi}}.
\label{eq3}
\end{equation} 
So the Maxwell-Boltzmann velocity distribution about the minimum velocity $\eta(v_{\rm min})$ can be given by
\begin{eqnarray}
  \eta(v_{\rm min})=\int_{v_{\rm min}} \frac{d^3 v}{v} f_\chi(v) \Theta (v-v_{\rm min}),
\end{eqnarray}
with
\begin{eqnarray}
 f_\chi(\vec{v}_\chi) \propto e^{-\frac{|\vec{v}_\chi +\vec{v}_E|^2}{v^2_0}} \Theta (v_{\rm esc}-|\vec{v}_{\chi}+\vec{v}_E|),
\end{eqnarray}
in which, we assume that the circular velocity $v_0=220\rm km/s$, cut at the escaping velocity
$v_{\rm esc} \sim 544 \rm km/s$ and the averaged Earth relative velocity $v_E=232$ km/s \cite{Smith:2006ym,Dehnen:1997cq}.
  
  Finally, we discuss the scattering cross section $\sigma_{e}$ between dark matter and electron from Eq. \eqref{rion}. In order to study the effect of momentum transfer, the reduced cross section $\bar{\sigma}_{e}$ and momentum dependent factor $F_{\rm DM}(q)$ are introduced. The scattering cross section can be written as
  \begin{equation}
  \sigma_{e}=\bar{\sigma}_{e}\vert F_{\rm DM}\vert^{2},
  \label{sigma}
  \end{equation}
  in which $\bar{\sigma}_{e}$ is a reduced cross section with momentum independence. It is given by
  \begin{equation}
      \bar{\sigma}_{e}=\mu_{\chi e}^{2}\frac{\overline{\vert M(q=q_{0})\vert^{2}}}{16\pi m_{\chi}^{2}m_{e}^{2}},
  \end{equation}
Then the form factor
$F_{\rm DM}(q)$ is given by \cite{Feldstein:2009tr}
\begin{equation}
\vert F_{\rm DM}(q)\vert^{2}=\frac{\overline{\vert M (q)\vert^{2}}}{\vert M(q=q_{0})\vert^{2}},
\label{fdmm}
\end{equation}
where $q_{0}=\alpha m_{e}$ is reference momentum.
We can see that all effect of momentum transfer is absorbed in form factor from Eq.~\eqref{fdmm}. In Ref.~\cite{Essig:2017kqs}, the form factor is considered as 1 and $1/q^2$ for heavy and light mediator in SI scattering. However, in SD scattering, the form factor are more complicated and dependent on the 
momentum transfer due to the appearance of $\gamma_{5}$ in mediator. 
All the form factor about scalar and vector dark matter are discussed 
in the next subsection.

\begin{widetext}
\begin{center}
\begin{table}[htbp]\renewcommand\arraystretch{2.2}
\begin{tabular}{|c|c|c|c|c|c|}
		\hline 
 DM coupling & Electron coupling & $\vert F_{\rm DM}(q)\vert^{2}$ & Massive& Massless\\
\hline
		 $a g_{\chi}\phi^{2}$ & $a g_{e}\bar{e}\gamma_{5}e$ & $\vert F_{\rm DM}^{1}\vert^{2}=\frac{q^{2}(m_{a}^{2}+q_{0}^{2})^{2}}{q_{0}^{2}(m_{a}^{2}+q^{2})^{2}}$ &$\approx \frac{q^2}{q_{0}^{2}}$&$\approx \frac{q_0^{2}}{q^2}$\\ 
		\hline
		$g_{\chi}(\phi^{\dag}\partial_{\mu}\phi-\phi\partial_{\mu}\phi^{\dag})A^{\mu}$&$g_{e}\bar{e}\gamma_{\mu}\gamma_{5}e A^{\mu}$&$\vert F_{\rm DM}^{2}\vert^{2}=\frac{q^{2}(m_{A}^{2}+q_{0}^{2})^{2}}{q_{0}^{2}(m_{A}^{2}+q^{2})^{2}}$& $\approx \frac{q^2}{q_{0}^{2}}$&$\approx \frac{q_0^{2}}{q^2}$\\
		\hline
		 $a g_{\chi}B_{\mu}B^{\mu}$&$a g_{e}\bar{e}\gamma_{5}e$&$\vert F_{\rm DM}^{3}\vert^{2}=\frac{q^{2}(m_{a}^{2}+q_{0}^{2})^{2}(4m_{\chi}^{2}q^{2}+12m_{\chi}^{4}+q^{4})}{q_{0}^{2}(m_{a}^{2}+q^{2})^{2}(4m_{\chi}^{2}q_{0}^{2}+12m_{\chi}^{4}+q_{0}^{4})}$& $\approx \frac{q^2}{q_{0}^{2}}$ & $\approx \frac{q_0^{2}}{q^2}$\\
		 \hline
		$g_{\chi}B_{\nu}^{\dag}\partial_{\mu}B^{\nu}A^{\mu}$&$g_{e}\bar{e}\gamma_{\mu}\gamma_{5}e A^{\mu}$&$\vert F_{\rm DM}^{4}\vert^{2}=\frac{q^{2}(m_{A}^{2}+q_{0}^{2})^{2}(4m_{\chi}^{2}q^{2}+12m_{\chi}^{4}+q^{4})}{q_{0}^{2}(m_{A}^{2}+q^{2})^{2}(4m_{\chi}^{2}q_{0}^{2}+12m_{\chi}^{4}+q_{0}^{4})}$&$\approx\frac{q^{2}}{q_{0}^{2}}$ &1\\ 
		&&\hspace{2cm}$\times\frac{(m_{A}^{2}m_{\chi}-m_{e}q^{2})(2m_{A}^{2}m_{e}+m_{A}^{2}m_{\chi}+m_{e}q^{2})}{(m_{A}^{2}m_{\chi}-m_{e}q_{0}^{2})(2m_{A}^{2}m_{e}+m_{A}^{2}m_{\chi}+m_{e}q_{0}^{2})}$&& \\\hline
		$g_{\chi}B_{\mu}\partial^{\mu}B_{\nu}A^{\nu}$&$g_{e}\bar{e}\gamma_{\mu}\gamma_{5}e A^{\mu}$&\hspace{-2.5cm}$\vert F_{\rm DM}^{5}\vert^{2}=\frac{q^{2}(m_{A}^{2}+q_{0}^{2})^{2}}{q_{0}^{2}(m_{A}^{2}+q^{2})^{2}}$ &$\approx\frac{q^{2}}{q_{0}^{2}}$&$\approx\frac {q_{0}^{2}}{q^{2}}$\\
	
		&&
\hspace{2cm}$\times\frac{-16m_{e}^{2}m_{\chi}^{4}-2m_{\chi}^{2}q^{2}(m_{e}-m_{\chi})^{2}+m_{e}^{2}q^{4}+2m_{e}m_{\chi}q^{4}}{-16m_{e}^{2}m_{\chi}^{4}-2m_{\chi}^{2}q_{0}^{2}(m_{e}-m_{\chi})^{2}+m_{e}^{2}q_{0}^{4}+2m_{e}m_{\chi}q0^{4}}$
& &\\
		\hline
	\end{tabular}
\caption{Different operators which induce the spin-dependent scattering between the dark matter and electron, together with the corresponding Form factors
and spin factors.}\label{table111}
	\end{table}
\end{center}
\vspace{-0.7cm}
\end{widetext}

\subsection{Benchmark model of scalar and vector dark matter}\label{sec3}
In order to check the effect of momentum transfer in SD scattering, the scalar dark matter with scalar mediator are studied in our numerical calculations. The interaction can be simplified as
\begin{equation}
    \mathcal{L_{\rm int}}\supset a g_{\chi}\phi^{2}+a g_{e}\bar{e}\gamma_{5}e,
    \label{int}
\end{equation}
in which, $g_{\chi}$ and $g_{e}$ is coupling constant of dark matter and electron, respectively. The scalar dark matter and electron are characterized by $\phi$ and $e$, respectively. The scalar mediator is denoted by $a$. 
With Eq. \eqref{int}, we can obtain the momentum dependent form factor $F_{\rm DM}$ from Eq. \eqref{fdmm}
\begin{eqnarray}
    \left\vert F_{\rm DM}\right\vert^{2}=\frac{q^{2}(m_{a}^{2}+q_{0}^{2})^{2}}{q_{0}^{2}(m_{a}^{2}+q^{2})^{2}},
    \label{fdmq}
\end{eqnarray}
where $m_{a}$ $(m_{\chi})$ is mass of mediator (dark matter). $q_{0}$ and q is reference and momentum transfer, respectively. In order to fully reflect the effect of momentum transfer. The interaction of vector dark matter is also considered. We listed all the interaction \cite{Fan:2010gt} between scalar ($\phi$) and vector ($B$) dark matter and electron with pseudo-scalar ($a$) and axial-vector (A) mediator in Tab. \ref{table111}. To compare with SI scattering, the corresponding form factor and the approximation in massive and massless of form factor are also listed in table. We can see when the 

mass of mediator is massive or massless, the form factor approximately equal to $1$, $q^{2}$ or $1/q^{2}$, these limit results are very simpleness and similarity. But in fact, the form factors have much more complicate q dependence analytically From the Tab. \ref{table111}.

These form factors are plotted in the left of Fig.~\ref{fdm} to explore effect of momentum transfer on SD scattering. For comparison, we also plotted the SI form factor with scalar or vector mediator for the scalar dark matter in right panel. 
The effect of momentum transfer is discussed in next section.

\section{Recoil spectra and Numerical results}\label{sec3}
\subsection{the procedure of numerical simulations}
We follow the procedure described in Ref.~\cite{Essig:2012yx} to calculate the constrains in experimental data. The total quantum number which can be produced by electron recoil energy $E_{\rm R}$ induce the observed electron $n_{e}$ and scintillation photons $n_{\gamma}$. $n^{(1)} = {\rm Floor}~(E_{\rm R}/W)$, in which, the $W=13.8$ eV is the minimum energy to produce a quantum number. We choose $f_{R}=0$, $f_{e}=0.83$ as the fiducial values.
The uncertainty is defined as $0<f_{R}<0.2$ , $12.4<W<16$ eV, and $0.62<f_{e}<0.91$. In our work we mainly consider the shells of electron is $(5s,~4d,~4p,~4s)$ which have energies $(13.3,~63.2,~87.9,~201.4)$ eV. There are also many additional quanta numbers in these shells under the effect of photoionization. The additional quanta numbers in shell $(5s,~4d,~4p,~4s)$ are $(0,~4,~6-10,~3-15)$, respectively \cite{Essig:2017kqs}. These quanta numbers also include the observed electron $n_{e}$ and scintillation photons $n_{\gamma}$. All quanta number is $(n^{1}+n^{2})$ which obey a binomial distribution. We can get the number of electron $n_{e}$ by calculating the corresponding distributions. These additional quanta number, binding energy of electron and photon energy are listed in the Tab. \ref{tabqe}
 \cite{Essig:2017kqs,Cao:2020bwd}.
\begin{table}[]
	\begin{tabular}{| l || l | l | l | l | l |}
		\hline 
		shell& 5$p^{6}$ & 5$s^{2}$ & 4$d^{10}$ & 4$p^{6}$ & 4 $s^{2}$\\
		\cline{1-6}
		Binding Energy[eV] & 12.6 & 25.7 & 75.6 & 163.5 & 213.8\\	
		\cline{1-6}
		Photon Energy[ev] &\quad-& 13.3 & 63.2 & 87.9 & 201.4\\
		\cline{1-6}
		Additional Quanta & 0 & 0 & 4 & 6-10 & 3-15\\
		\cline{1-6}
		\hline
	\end{tabular}
\caption{The number of additional quanta and binding energy 
from the Xenon ($5p^{6},~5s^{2},~4d^{10},~4p^{6},~4s^{2}$) shells. }\label{tabqe}
\end{table}
\begin{figure}[H]
	\centering
\scalebox{0.56}{\epsfig{file=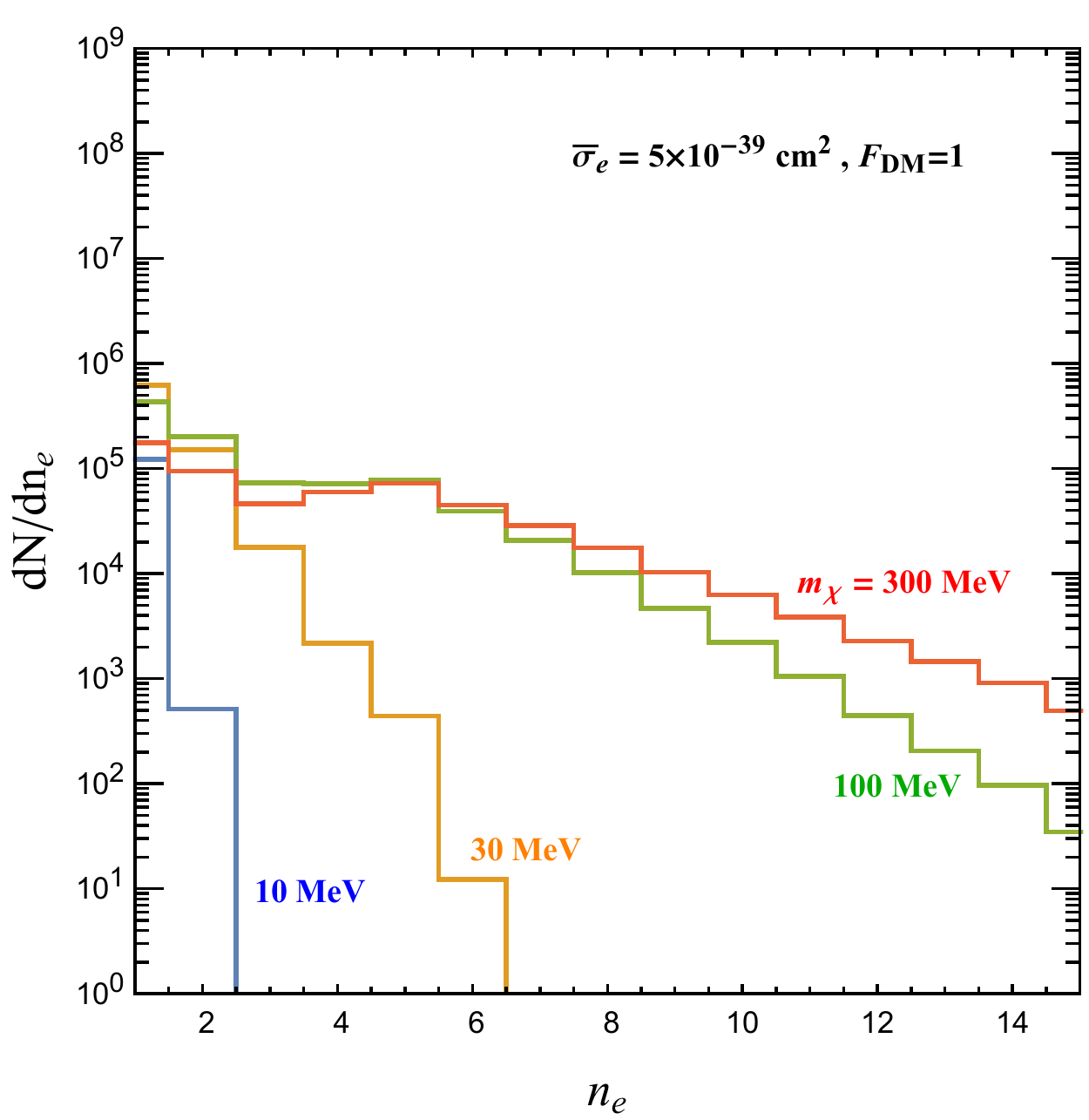}}
\scalebox{0.56}{\epsfig{file=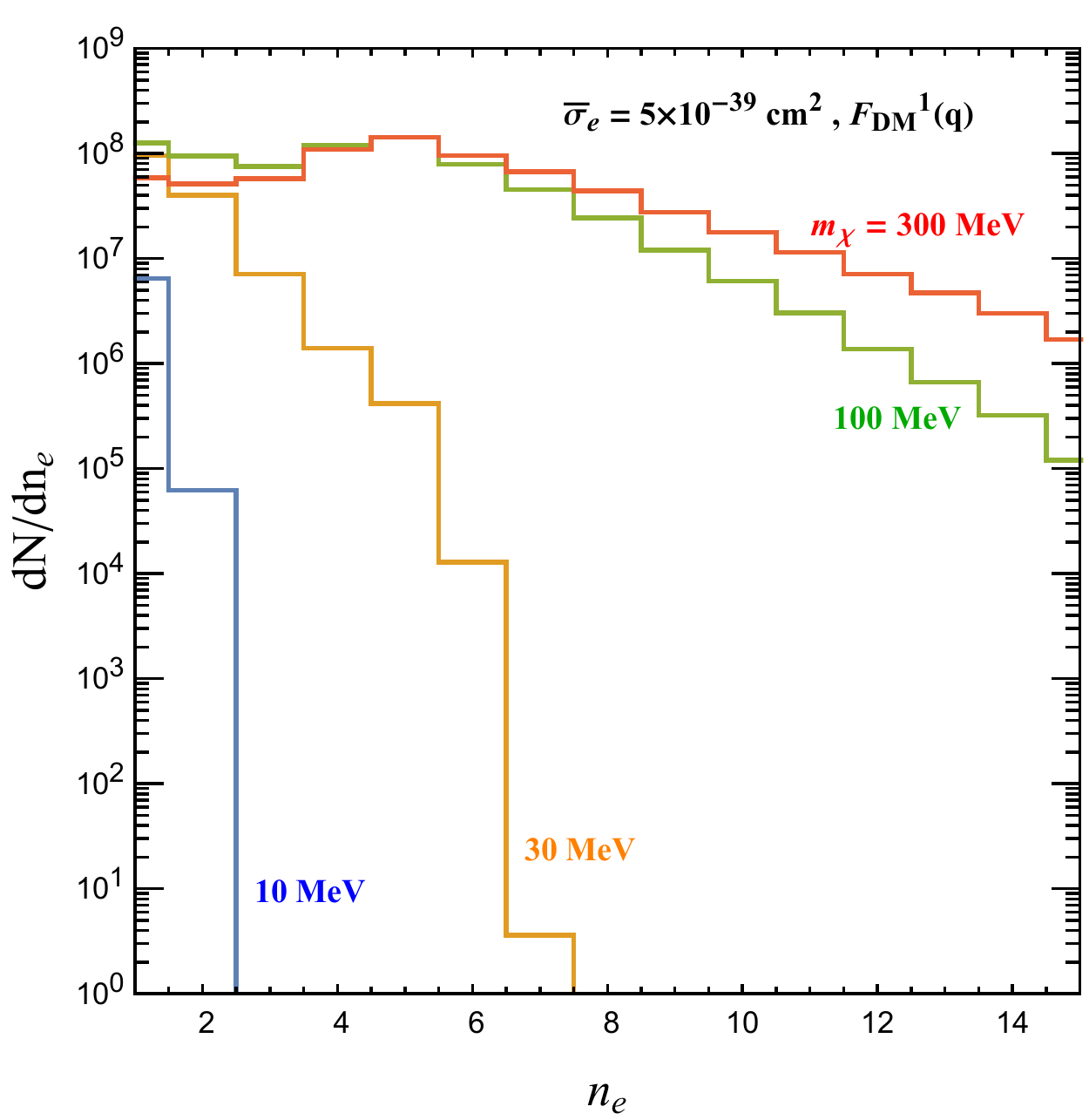}}
\caption{The spectra of electron from dark matter scattering with electron in $F_{\rm DM}=1$ (top) and $F_{\rm DM}^{1}(q)$ (bottom). We choose $m_{\chi} = 10$ MeV (blue), 30 MeV (orange), 100 MeV (green) and 300 MeV (red) for the comparison. And the fiducial cross section and exposure are chosen as $\bar{\sigma}_{e}=5 \times 10^{-39} \rm cm^{2}$ and 1000 kg-yr.}
	\label{peevent}
\end{figure}

On the experimental side, the weak signal of electron 
can not be observed. In order to measure the amount electron $n_{e}$, we must amplify the electron signal by photomultipier tuber. The signal of electron is converted to the photoelectrons (PE).
We apply the experiment data from XENON10 data \cite{Essig:2012yx} (15 kg-days), XENON100 data \cite{XENON100:2011cza,XENON:2016jmt} (30 kg-years) and 
XENON1T data \cite{XENON:2019gfn} (1.5 tones-years) to compare our theory result. The boundary limits of SD scattering cross section 
are obtained by comparing these result. 
The number of PE and $n_{e}$ obey Gaussian distribution. The PE can be produced by calculating this Gaussian function which has mean $n_{e}\mu$ and width $\sqrt{n_{e}\sigma}$, in which $\mu = 27~(19.7, 11.4)$ and $\sigma = 6.7~ (6.2, 2.8)$ for XENON10 (XENON100, XENON1T). We follow the Ref. \cite{Essig:2012yx,XENON100:2011cza,XENON:2019gfn} to get the PE bins. Note that, in our numerical evaluation the range of PE bin in the XENON1T is only chosen as [165,~275].  

For SI scattering, the form factor will be suppressed by the momentum transfer in light mediator from literature \cite{Essig:2012yx}. While for SD interaction, the form factors are more complicated from Eq. \eqref{fdmq} and Tab. \ref{table111}. To explore the effect of momentum transfer on SD scattering, the spectra of electron from $F_{\rm DM}=1$ (top) and $F_{\rm DM}^{1}(q)$ (bottom) are plotted in the Fig. \ref{peevent}. In which, 
the mediator $m_{a}$ is set as 0.5 MeV and the fiducial cross section $\bar{\sigma}_{e}$ is set as $5\times 10^{-39}$ $\rm{cm}^2$. We choose the mass of dark matter are $m_{\chi}=10$ MeV, 30 MeV, 100MeV, 300MeV for comparison, and the exposure is set as 1000 kg-yr. From Fig. \ref{peevent}, it can be seen that the number of electron is same in the same dark matter mass. This is because the quantum number depend on recoil energy $E_{R}$ which can be produced by the incoming energy of dark matter ($1/2m_{\chi}v^{2}$). However, the event rates of the $F_{\rm DM}^{1}(q)$ (bottom) can always be enhanced two orders of magnitude greater than 
$F_{\rm DM}=1$ for all mass of dark matter. We understand this result from Eq. \eqref{rion}. It show that the events rates will depend on form factor $F_{\rm DM}$. This imply that the momentum transfer plays an important role in dark matter form factor.
\begin{figure}[htbp]
	\centering
\scalebox{0.37}{\epsfig{file=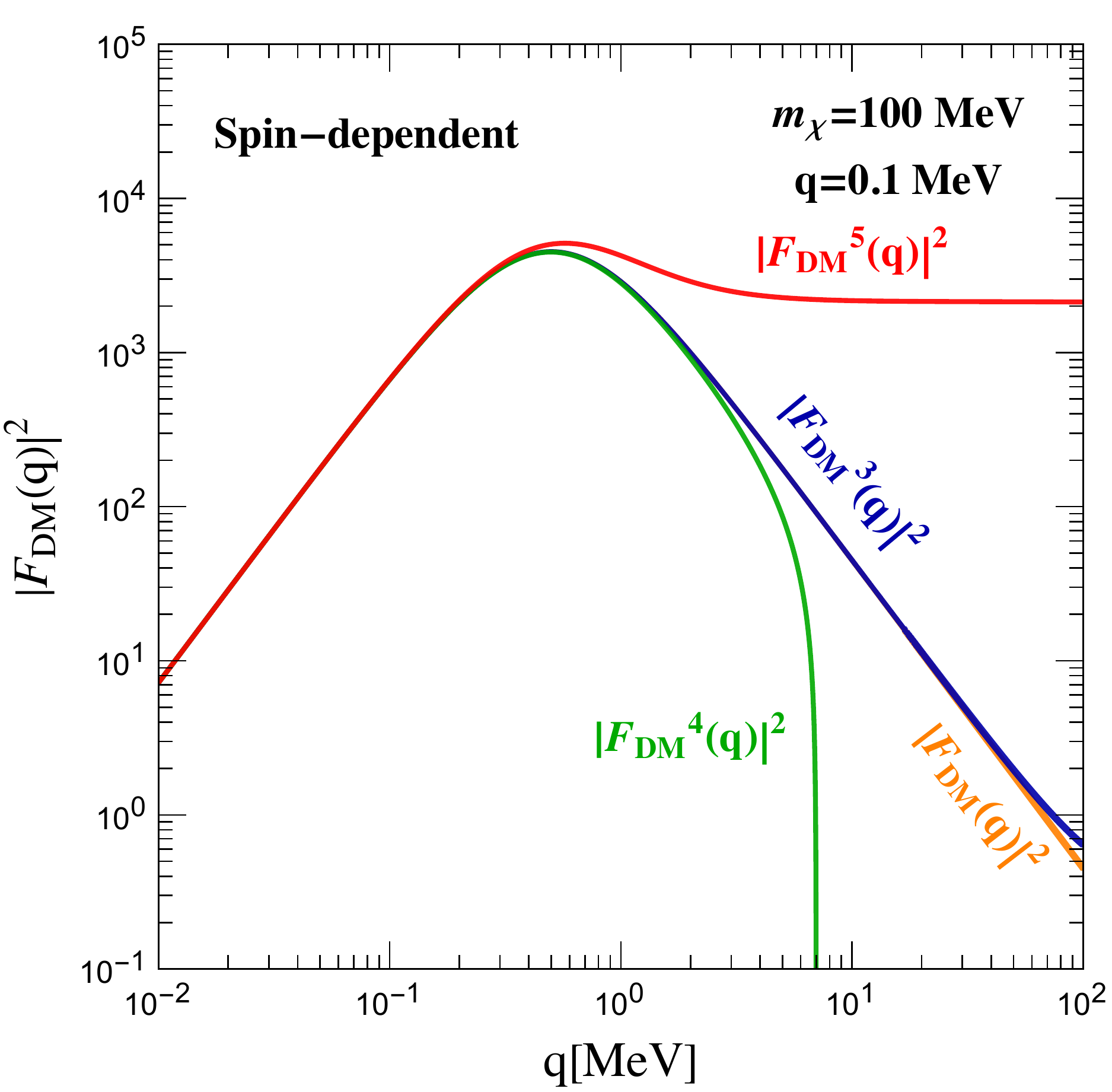}}
\scalebox{0.56}{\epsfig{file=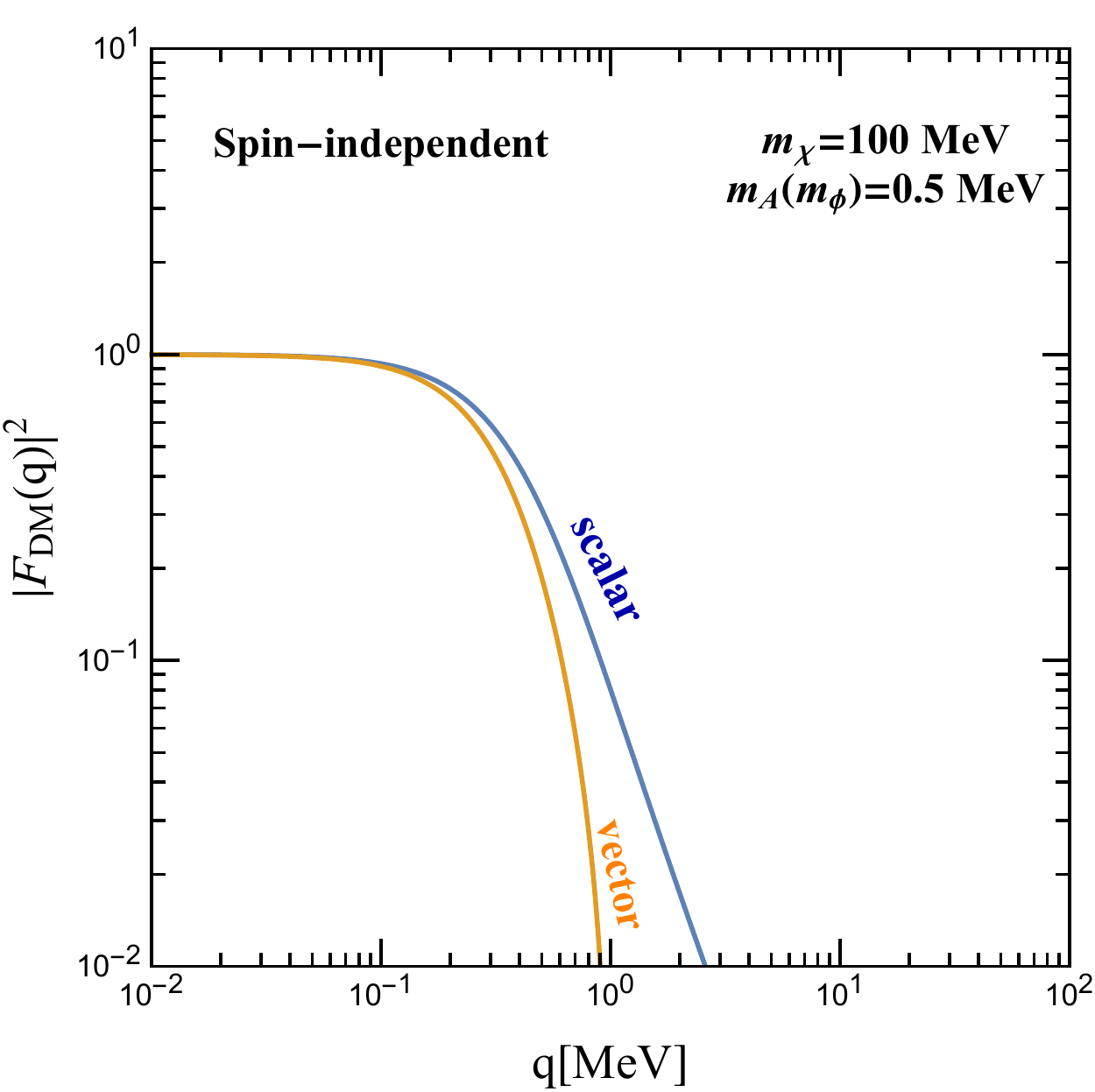}}
\caption{Left panel: the form factor $\vert F_{\rm DM}(q)\vert$ for the 
SD scattering of the Fermionic dark matter. 
Right panel: the form factor $\vert F_{\rm DM}(q)$
for the SI scattering with a scalar or vector mediator for the Fermionic dark matter. $m_{\chi}=100$MeV for each panel.}
	\label{fdm}
\end{figure}
\begin{figure}[htbp]
	\centering
	\scalebox{0.38}{\epsfig{file=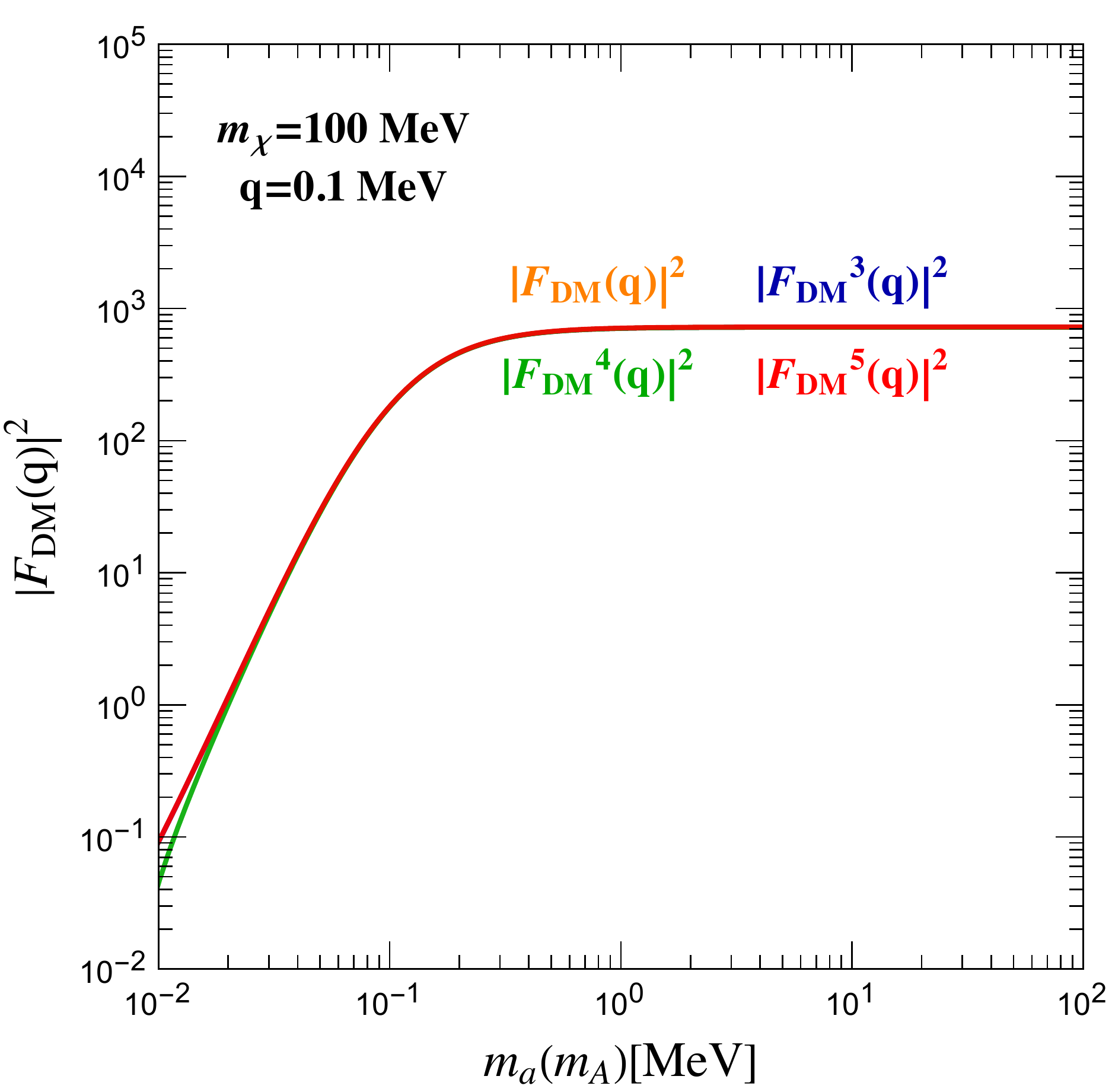}}
\caption{The form factor $\vert F_{\rm DM}(q)\vert$ in different mass of mediator are shown, in which, the momentum transfer $q$ is set as 0.1 MeV and the mass of dark matter is chosen as 100 MeV.}
	\label{fdm2}
\end{figure}

\begin{figure*}[htbp]
	\centering
	\scalebox{0.48}{\epsfig{file=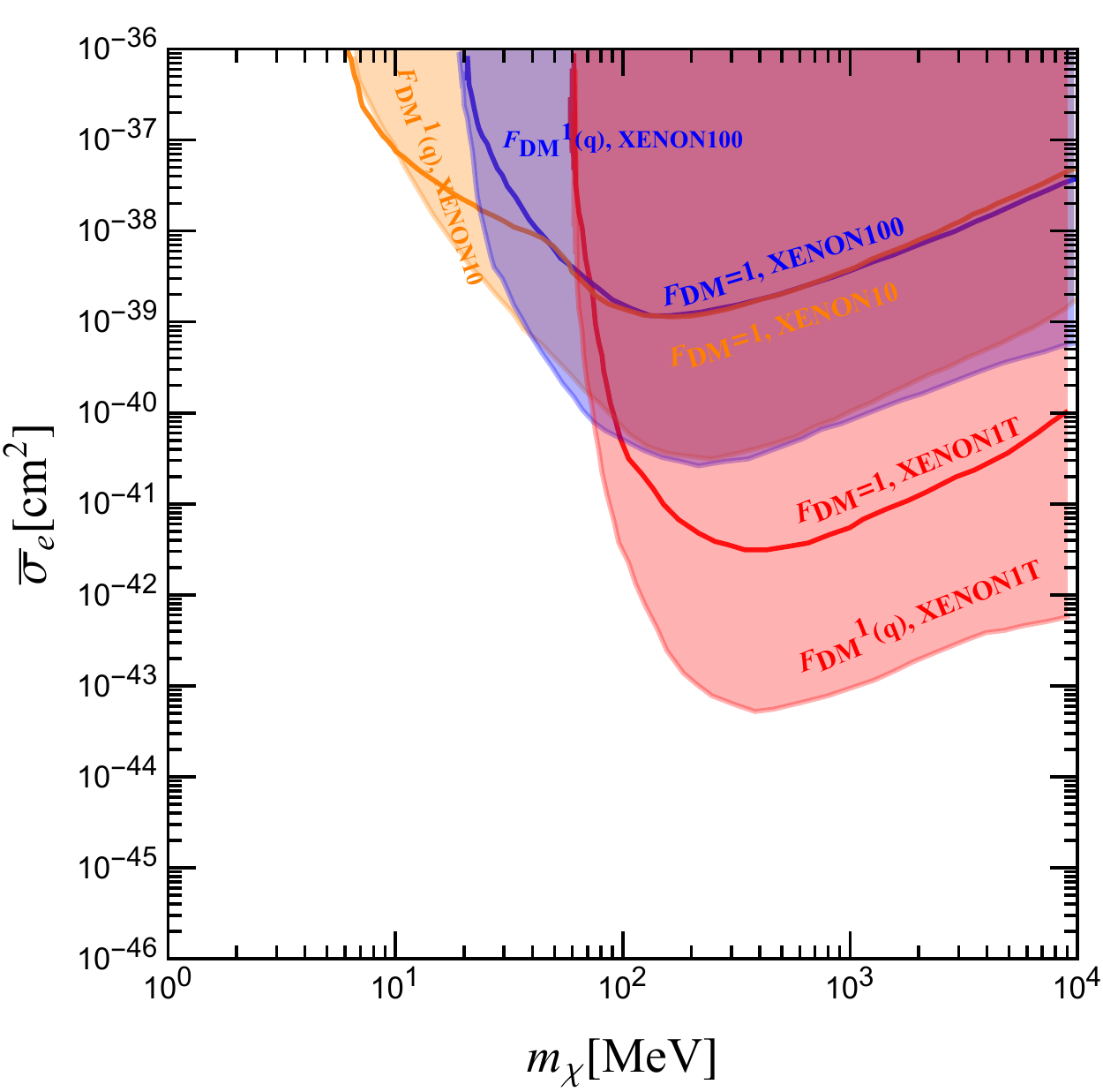}}
	\scalebox{0.48}{\epsfig{file=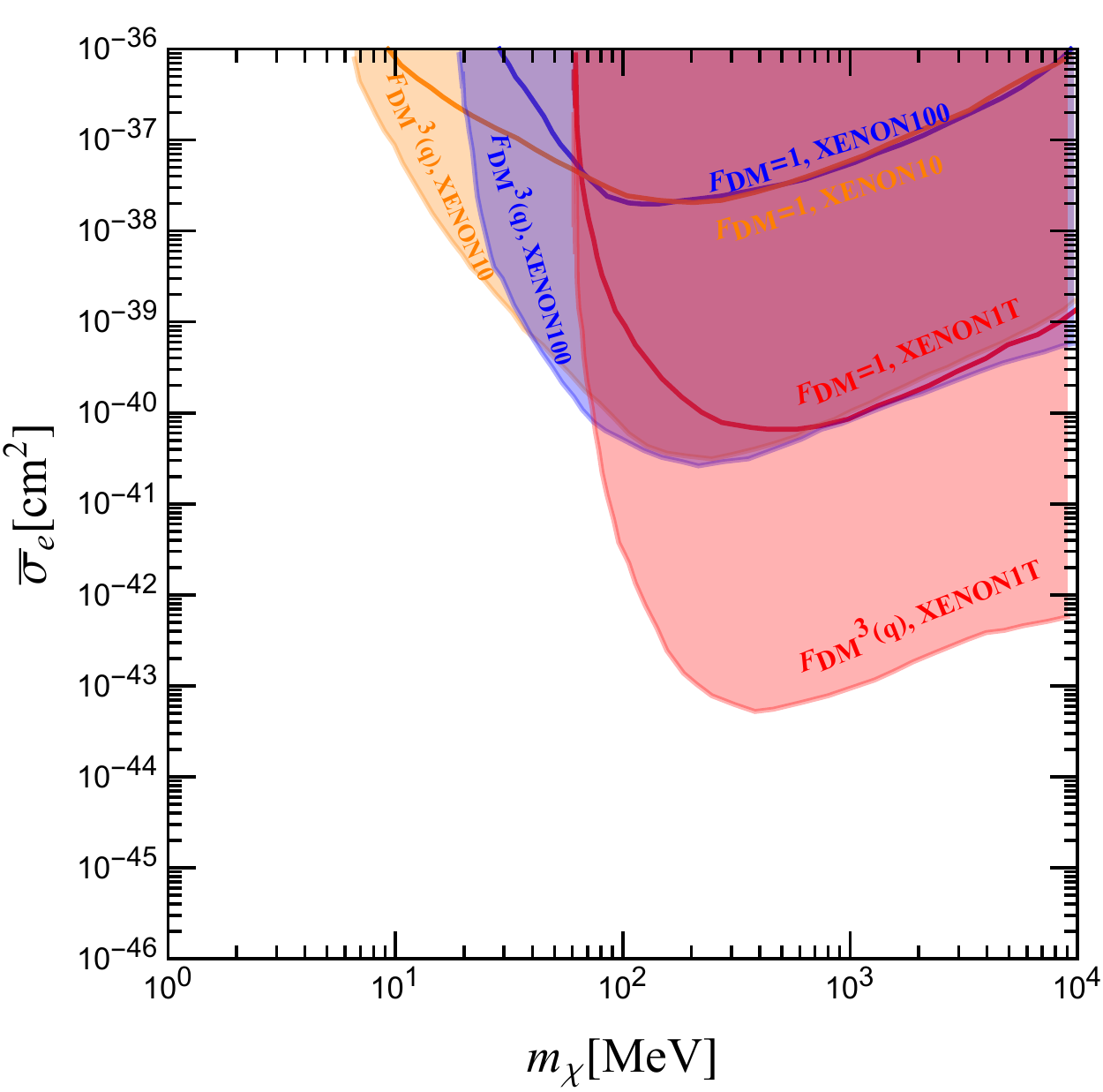}}
	\scalebox{0.48}{\epsfig{file=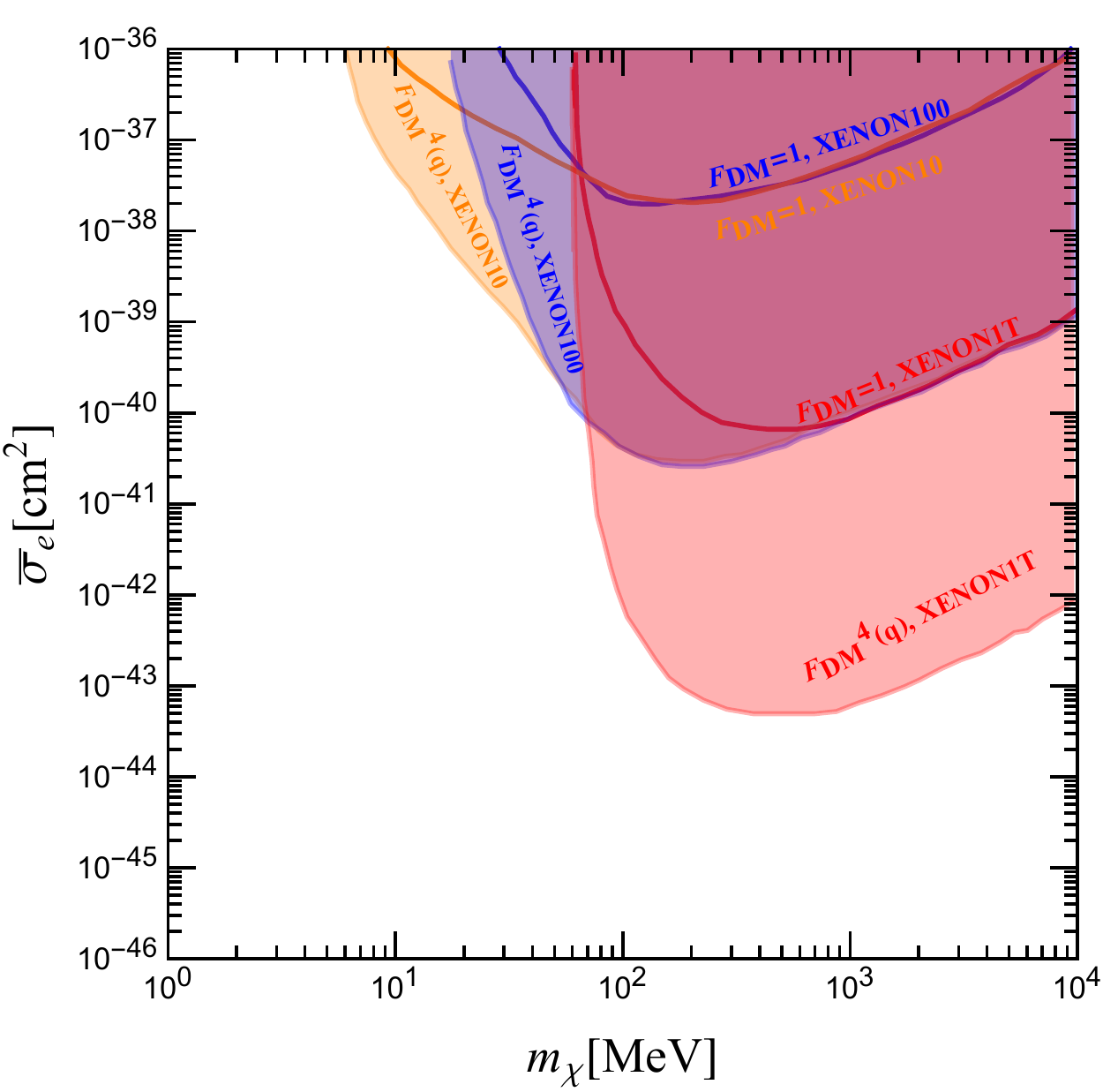}}
	\scalebox{0.48}{\epsfig{file=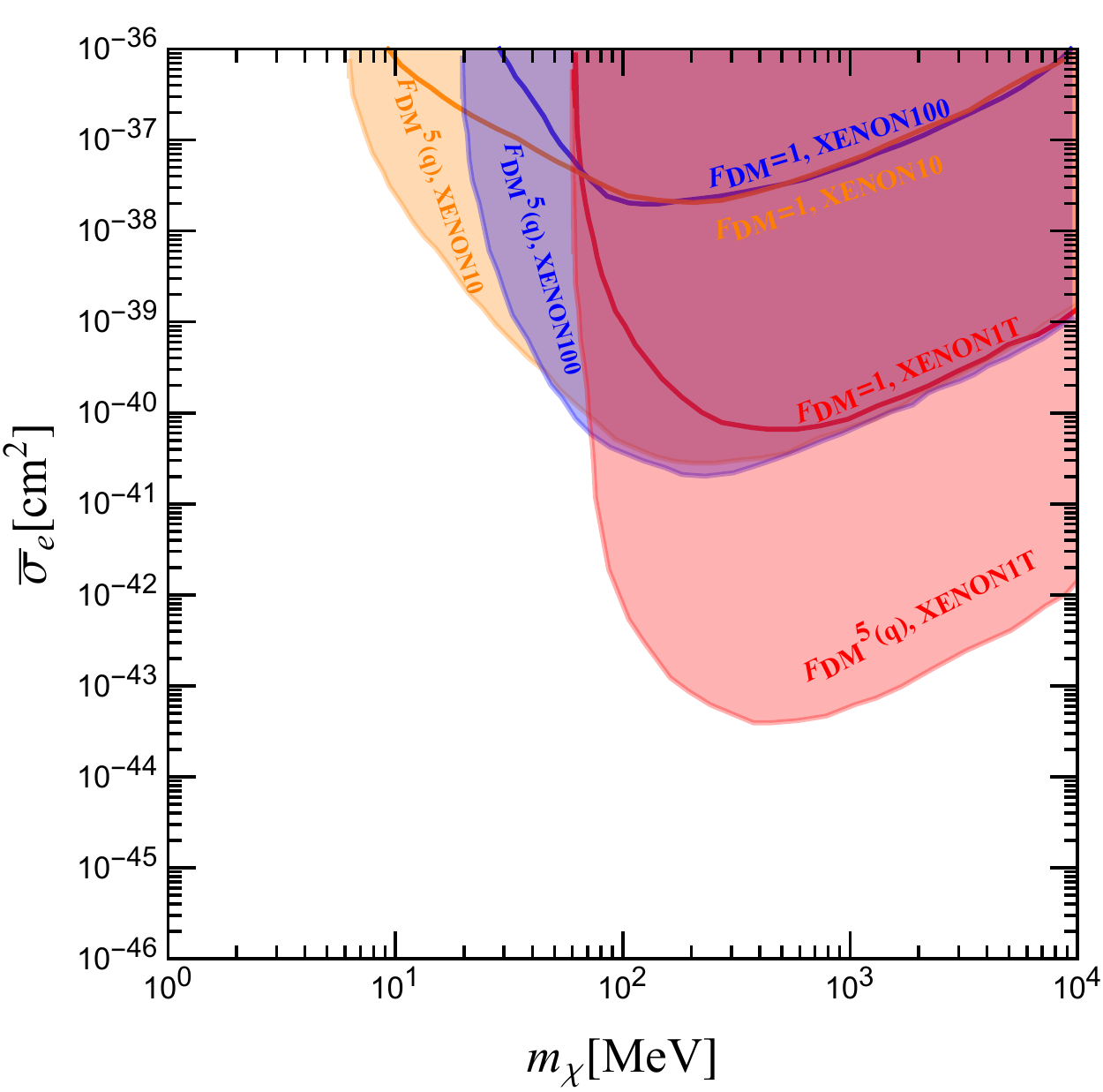}}
\caption{The limits of the cross sections got from Fermionic DM-electron 
scattering are shown in the top panel for $F_{\rm DM}^{1}$ 
and the bottom panel for $F_{\rm DM}^{3}$.
The corresponding results from $F_{\rm DM}=1$ (line) are also shown
in each panel for the comparison. 
The orange shades (lines) show the results got from XENON10 data,
the blue ones show results of XENON100 data and the red ones shows the results
of XENON1T data, respectively.}	\label{sigmae}			
\end{figure*}
The dependence of SD form factor on momentum transfer are shown in left panel of the Fig. \ref{fdm}. In which, all the form factors from Tab. \ref{table111} are considered. The mass of the dark matter is set as $m_{\chi}= 100 \rm MeV$ and the mass of mediator is chosen as $m_{A}(m_{\phi})=0.5\rm MeV$. In order to explore the difference between SD scattering and SI scattering in form factor with light mediator, we also plotted the form factor of SI in the right panel. We can see that most form factors of SD scattering are greater than SI scattering. And the maximum value of SD form factor can be four orders of magnitude stronger than $F_{\rm DM}=1$. The SD form factor can be the maximum value when the momentum transfer $q$ is approximately equal to mediator mass. For different interaction, the form factor can be difference, so the form factor is model-dependent. We find that the mass of mediator also effect the value of form factor in Eq. \eqref{fdmq}. Therefore we plotted the dependence of SD form factor on the mass of mediator in Fig. \ref{fdm2}. We can see the form factor will not change when $m_{a}(m_{A})$ is greater than 0.1 MeV in Fig. \ref{fdm2}. However, for light mediator, the form factor will be suppressed by mass of mediator. So the mediator also play an important role in form factor. Therefore, we set the mass of mediator as 0.5 MeV to show the effect of momentum transfer.

\subsection{The result and analysis of numerical simulations}

  As we discussed above, In light mediator the SD and SI form factor can be suppressed by momentum transfer. This result satisfies the long-range interaction. And it can also be enhanced by the mass of mediator. However, in heavy mediator, the SI form factor will be constant 1 which is called contact interaction. The SD form factor can be enhanced by momentum transfer. The reason for this result is that the SD and SI scattering are $p$-wave scattering and $s$-wave scattering, respectively. Besides, when we define the reduced cross section $\bar {\sigma}_{e}$ in specific momentum $q_{0}$, the corresponding form factors are always proportional to the momentum transfer. The different $q_{0}$ can make the reduced cross section $\bar{\sigma}_{e}$ and form factor change greatly. Thus the definition points $q_{0}$ are also more important in the SD scattering. The specific form factor $F_{\rm DM}(q)$ can be seen from the Tab. \ref{table111} and the numerical results in the Fig. \ref{fdm}.
  
In order to explore the limits of SD scattering, we plotted the dependence of SD reduced cross section $\bar{\sigma}_{e}$ on mass of dark matter in Fig. \ref{sigmae}. From Eq. \eqref{rion}, we can see that the reduced cross section $\bar{\sigma}_{e}$ will decrease as form factor $F_{\rm DM}$ increase when the counting rate $R$ is a constant value. 
In Fig. \ref{sigmae}, we use the dotted line and shades to show the result of $F_{\rm DM}=1$ and $F_{\rm DM}(q)$, respectively. The orange area, blue area and red area denote XENON10 data, XENON100 data and XENON1T data, respectively. The top picture of Fig. \ref{sigmae} is the result of the pseudo-scalar mediator. The upper left represent scalar dark matter $(F_{\rm DM}^{1}(q))$ and the right one is vector dark matter$(F_{\rm DM}^{3}(q))$. The bottom picture of Fig. \ref{sigmae} is the result of vector dark matter in axial-vector mediator ($F_{\rm DM}^{4}(q)$, $F_{\rm DM}^{5}(q)$). Comparing with $F_{\rm DM}=1$ (dotted line), we can see that the cross section on SD scattering will be significantly enhanced. This is because the form factor can be enhanced as momentum transfer improve. The improved form factor can affect the reduced cross section from Eq. \eqref{sigma}. We also find the exclusion limits of SD scattering cross section are more sensitive in the region of $m_{\chi}>100\rm MeV$. It can be understood from the Fig. \ref{peevent}, in which the range of electron $n_{e}$ is very small in light dark matter ($m_\chi <100$MeV). When $m_{\chi}>100 \rm MeV$ the range of electron and the event can be the largest, but the range of electron is not change and the event can be reduced when $m_{\chi}>300\rm MeV$.

In above calculation, we set the mass of mediator $m_{a}$ $(m_{A})$ as 0.5 MeV. This is because in this case, the form factor can be get the maximum value. In which, the momentum transfer $q$ is approximately equal to mass of mediator from Fig. \ref{fdm}. We also find that the form factor will dependent on the mass of mediator from Fig .\ref{fdm2}. The definition of $q_{0}$ and mass of mediator can affect the form factor and reduced cross section. Thus we show all the exclusion limits in $q_{0}=\alpha m_{e}$ and $m_{A}(m_{a})=0.5~\rm MeV$ for the comparisons between the different models. The results are plotted in Fig. \ref{sigmae}. Furthermore we can get following conclusions
\begin{enumerate}
\item The $q_{0}$ must be defined at a specific value $\alpha m_{e}$ to compare the reduced cross section $\bar{\sigma}_{e}$ of SD and SI scattering. In this scene, the reduced cross section of SD scattering can be three orders of magnitude large than SI scattering due to the effect of transferred momentum.

\item The form factor $F_{\rm DM}^{1}(q)$, $F_{\rm DM}^{2}(q)$, $F_{\rm DM}^{4}(q)$ and $F_{\rm DM}^{5}(q)$ can recover $\Lambda^2/q^2$ and the $F_{\rm DM}^{3}$ can recover 1 in massless mediator. From this result, we can know the from factors of SD can recover the SI in vanishing mediator mass. All the form factor will proportional to $q^{2}$ in the heavy dark matter which can enhance the exclusion limits.
\end{enumerate}

\section{Conclusion}\label{sec4}

The experiment efforts to search for dark matter-electron coupling motivate consideration of scalar and vector dark matter candidate. In this paper, we calculate all the form factor from dark matter-electron couplings mediated by pseudo-scalar and axial-vector. We also plotted the dependence on mass of dark matter of SI and SD reduced cross section.

Our result show that the form factor of SD scattering can be four orders of magnitude larger than SI scattering when the momentum transfer is less than the mass of mediator $(q<0.5~\rm MeV)$. For light mediator $(q>0.5~\rm MeV)$, the form factor will be suppressed in SI and SD scattering. Due to the peak value of SD scattering is much greater than 1, so the SD scattering form factor will greater than SI scattering. The exclusion limits of SD scattering cross section which is dependent on form factor will also be enhanced in light mediator. It can be $10^{-44}$ $\rm cm^{2}$ in XENON1T. This result is same as fermion dark matter from literature \cite{Liu:2021avx}. The reason for this result is that the SD scattering are p-wave scattering scattering. We also find that the form factor is not affected by the mediator when the mass of mediator is greater than momentum transfer. However, when $m_{a}<0.1$ MeV, the form factor will be improved as mass of mediator increases. 
\vspace{3mm}
\section{Acknowledgement}\label{sec5}
This work was supported by the Natural Science Foundation of China under grant number 11775012. We would like to thank Bin Zhu and Wulong Xu for many useful discussions on SD theory and for procedural guidance. The author also is grateful to Lei Wu, Liangliang Su and Aichen Li for many useful discussions on form factor and the effect of transferring momentum.


\begin{widetext}
	\section*{Appendix: Calculation of the Response function}

Ref. \cite{Liu:2021avx} apply atom many-body method to calculate SI and SD response function, 
we follow this method to deal with the relationship between SI and SD response function. 
Firstly, for SI response function ${\cal R}_{\rm SI}^{\rm ion}(E_R,q)$
\begin{equation}
 {\cal R}_{\rm SI}^{\rm ion}(E_R,q)=\sum_{J_{f}}\overline{\sum_{J_{i}}}\vert\langle J_{f}\vert\sum_{i=1}^{Z}e^{i{\bm q}\cdot{\bm r}_{i}}\vert  J_{i}\rangle\vert^{2}
 \delta(E_{J_{f}}-E_{J_{i}}-E_R),
\end{equation}
in which $\vert J_{i}\rangle=\vert L_{i}J_{j_{i}}M_{i}\rangle$ 
and $\vert J_{f}\rangle=\vert L_{f}J_{j_{f}}M_{f}\rangle$ label initial and final states, 
respectively. Where $L_{i}$ and $L_{f}$ are other quantum numbers. 
The $M_{i}$ and $M_{f}$ can be reduced by Wigner-Eckart theorem. And delta function label 
the relation of initial and final energy: $E_{J_{f}}=E_{J_{i}}+E_R$. SI response function 
can be written the following equation from literature \cite{Fitzpatrick:2012ix, Liu:2021avx}
\begin{equation}
    {\cal R}_{\rm SI}^{\rm ion}(E_R,q)=\sum_{J_{f}}\sum_{J_{i}}\sum_{J=0}\frac{4\pi}{2J+1}\vert \langle J_{f}\vert \sum_{i=1}^{Z}j_{J}(qr_{i})Y_{J}^{M_{J}}(\Omega_{r_{i}})\vert J_{i}\rangle \vert^{2}\delta(E_{J_{f}}-E_{J_{i}}-E_R),
    \label{ymj}
\end{equation}
where $Y_{J}^{M_{J}}$ is spherical harmonics and $j_{J}(qr_{i})$ is the spherical 
Bessel function. $\vert J_{f}\rangle$ and $\vert J_{i}\rangle$ are many-body, 
but the operators are one-body, so the multipole operator $\hat{O}_{J_{f}J_{i}}$ 
can be represented by the sum of one-body$\sum_{i=1}^{Z}\hat{o}_{J_{f}J_{i}}(i)$. 
It can be written as $\sum_{i=1}^{Z}Y_{JM}(\Omega_{r_{i}})=Y_{JM}(\Omega_{r})$. So the Eq. \ref{ymj} can be simplified by applying the following equation
\begin{equation}
    \langle J_{f}\vert \hat{O}_{J_{f}J_{i}}\vert\ J_{i}\rangle=\sum_{\alpha \beta}\psi_{J_{f}J_{i}}(\alpha, \beta)\langle \alpha\vert \hat{o}_{J_{f}J_{i}}\vert\beta\rangle,
\end{equation}
where $\psi_{J_{f}J_{i}}(\alpha, \beta)$ that is one-body density matrix can be defined as the following equation 
\begin{equation}
    \psi_{J_{f}J_{i}}(\alpha, \beta)=\frac{1}{\sqrt{2J+1}}\langle J_{f}\vert [\hat{c}_{\alpha}^{\dagger}\otimes\hat{c}_{\beta}]\vert J_{i}\rangle,
    \label{RSI}
\end{equation}
in which, the $\hat{c}_{\alpha}^{\dagger}$ and $\hat{c}_{\beta}$ are creation and 
annihilation operators of electron. For single-particle basis
$\vert \alpha\rangle=\vert n_{\alpha}(l_{\alpha}1/2)j_{\alpha}m_{\alpha}\rangle$, 
$\vert \beta\rangle=\vert n_{\beta}(l_{\beta}1/2)j_{\beta}m_{\beta}\rangle$ 
has the general coordinate-space form
\begin{equation}
    \vert \alpha\rangle=\vert n_{\alpha}(l_{\alpha}1/2)j_{\alpha}m_{\alpha}\rangle={\cal R}_{n_{\alpha}l_{\alpha}j_{\alpha}}(x)[Y_{l_{\alpha}}(\Omega_{x})\otimes\chi_{1/2}],
\end{equation}
in which $\chi_{1/2}$ is the Pauli spin, ${\cal R}_{n_{\alpha}l_{\alpha}j_{\alpha}(x)}$ 
is radial wave function. Then we use Wigner 3-j, 6-j 
and 9-j symbols \cite{Liu:2021avx, Haxton:2008zza} to reduce Eq. (\ref{RSI}), 
it can be reduced to the following equation
    \begin{equation}
    \begin{split}
    {\cal R}_{\rm SI}^{\rm ion}(E_{R},q)&=\sum_{n_{\beta}l_{\beta}j_{\beta}}\sum_{n_{\alpha}l_{\alpha}j_{\alpha}}\sum_{L=0}4\pi\vert\langle n_{\beta}(l_{\beta}1/2)j_{\beta}\vert j_{L}(qr)Y_{L}(\Omega_{r})\vert n_{\alpha}(l_{\alpha}1/2)j_{\alpha}\rangle\vert^{2}\delta(E_{J_{f}}-E_{J_{i}}-E_{R})\\
&=2\sum_{n_{\beta}l_{\beta}}\sum_{n_{\alpha}l_{\alpha}}\sum_{L=0}(2l_{\alpha}+1)(2l_{\beta}+1)(2L+1)\begin{pmatrix}
l_{\alpha} & L & l_{\beta}\\ 0 & 0 & 0
\end{pmatrix}^{2}\vert\langle n_{\alpha}l_{\alpha}\vert j_{J}(qx)\vert n_{\beta}l_{\beta}\rangle\vert^{2}\delta(E_{J_{f}}-E_{J_{i}}-E_{R}).
    \end{split}
\end{equation}
Then for SD response function ${\cal R}_{\rm SD}^{\rm ion}(E_{R},q)$, it can be written as 
\begin{equation}
 {\cal R}_{\rm SD}^{\rm ion}(E_{R},q)=\sum_{J_{f}}\overline{\sum_{J_{i}}}\sum_{k}\vert\langle J_{f}\vert\sum_{i=1}^{Z}e^{i{\bm q}\cdot{\bm r}_{i}}\vec{\boldsymbol{\sigma}}^{s}_{ik}\vert  J_{i}\rangle\vert^{2} \delta(E_{J_{f}}-E_{J_{i}}-E_{R}).
\end{equation}
As the same way, we apply the atom many-body method from Ref \cite{Fitzpatrick:2012ix, Liu:2021avx}, the 
SD response function can be reduced as
\begin{equation}
    \begin{split}
 {\cal R}_{\rm SD}^{\rm ion}(E_{R},q)=&\sum_{J_{f}}\overline{\sum_{J_{i}}}\sum_{k}\frac{4\pi}{2J+1}\vert\langle J_{f}\vert\sum_{i=1}^{Z}\boldsymbol{M}_{JL}\cdot\boldsymbol{\sigma}^{s}\vert  J_{i}\rangle\vert^{2}\delta(E_{J_{f}}-E_{J_{i}}-E_{R}),
 \end{split}
\end{equation}
in which, $\boldsymbol{\sigma}^{s}
=\gamma^{0}\boldsymbol{\gamma}=\begin{pmatrix}\boldsymbol{\sigma} & 0 \\ 0 & \boldsymbol{\sigma} 
\end{pmatrix}$ is spin operator of electron. The term $e^{i\boldsymbol{q}\boldsymbol{r_{i}}}\vec{\boldsymbol{\sigma}}$ satisfy the following equation 
\begin{equation}
   e^{i\boldsymbol{q}\boldsymbol{r_{i}}}\vec{\boldsymbol{\sigma}}=\boldsymbol{M}_{JL}\cdot\boldsymbol{\sigma}^{s}= j_{L}(qx)\boldsymbol{Y}_{JL1}^{M_{J}}(\Omega_{x})\cdot\boldsymbol{\sigma}^{s},
\end{equation}
\begin{equation}
    \boldsymbol{Y}_{JL1}^{M_{J}}(\Omega_{x})=[Y_{L}\otimes\boldsymbol{e_{1}}]_{JM_{J}}=\sum_{m\lambda}\langle Lm1\lambda\vert(L1)JM_{J}\rangle Y_{L}^{m}(\Omega_{x})\boldsymbol{e}_{1\lambda}.
\end{equation}
The spherical harmonics $Y_{J}^{M_{J}}$ and the unit vector $\boldsymbol{e_{1}}$ can form vector 
spherical harmonics $\boldsymbol{Y}_{JL1}^{M_{J}}(\Omega_{x})$. We can get the following 
result by using Winger 3-j, 6-j and 9-j \cite{Liu:2021avx, Haxton:2008zza} for single-particle state.
\begin{equation}
    \begin{split}
    {\cal R}_{\rm SD}^{\rm ion}(E_{R},q)&=\sum_{n_{\beta}l_{\beta}j_{\beta}}\sum_{n_{\alpha}l_{\alpha}j_{\alpha}}\sum_{L=0}4\pi\vert\langle n_{\beta}(l_{\beta}1/2)j_{\beta}\vert j_{L}(qr)\boldsymbol{Y}_{JL_{i}}(\Omega_{x})\cdot \vec{\boldsymbol{\sigma}}^{s}\vert n_{\alpha}(l_{\alpha}1/2)j_{\alpha}\rangle\vert^{2}\delta(E_{J_{f}}-E_{J_{i}}-E_{R}) \\
    &=6\sum_{n_{\beta}l_{\beta}}\sum_{n_{\alpha}l_{\alpha}}\sum_{L=0}(2l_{\alpha}+1)(2l_{\beta}+1)(2j_{\alpha}+1)(2j_{\beta}+1)(2J+1)(2L+1)\begin{Bmatrix}
    l_{\alpha} & l_{\beta} & L\\ \frac{1}{2} & \frac{1}{2} & 1\\ j_{\alpha} & j_{\beta} & J\end{Bmatrix}^{2}\\
    &\times\begin{pmatrix}
    l_{\alpha} & L & l_{\beta}\\0 & 0 & 0\end{pmatrix}^{2}\vert\langle n_{\alpha}l_{\alpha}j_{\alpha}\vert j_{J}(qx)\vert n_{\beta}l_{\beta}j_{\beta}\rangle\vert^{2}\delta({E_{J_{f}}-E_{J_{i}}-E_{R}})\\
    &=2\sum_{n_{\beta}l_{\beta}}\sum_{n_{\alpha}l_{\alpha}}(2l_{\beta}+1)(2l_{\alpha}+1)\left\{3\begin{pmatrix}
    l_{\alpha} & 0 & l_{\beta}\\0 & 0 & 0\end{pmatrix}^{2}+\sum_{L=1}[(2L+1)+2(L-1)+1+2(L+1)+1]\begin{pmatrix}
    l_{\alpha} & L & l_{\beta}\\0 & 0 & 0\end{pmatrix}^{2}\right\}\\
    &\times \langle n_{\alpha}l_{\alpha}\vert j_{J}(qx)\vert n_{\beta}l_{\beta}\rangle_{NR}\delta({E_{J_{f}}-E_{J_{i}}-E_{R}})\\
    &=6\sum_{n_{\beta}l_{\beta}}\sum_{n_{\alpha}l_{\alpha}}\sum_{L=0}(2l_{\beta}+1)(2l_{\alpha}+1)(2L+1)\begin{pmatrix}
    l_{\alpha} & L & l_{\beta}\\0 & 0 & 0\end{pmatrix}^{2}\langle n_{\alpha}l_{\alpha}\vert j_{J}(qx)\vert n_{\beta}l_{\beta}\rangle_{NR}\delta({E_{J_{f}}-E_{J_{i}}-E_{R}})\\
    &=3{\cal R}_{\rm SI}^{\rm ion}(E_{R},q).
    \end{split}
 \end{equation}
\end{widetext}
\bibliography{refs}
\end{document}